\documentclass[12pt]{article}
\usepackage{graphicx}
\def\mydate{March 11, 2005}
\def\ignore#1{{}}

\newcounter{sxn}

\newcounter{axn}

\date{}

\newdimen\mybaselineskip
\renewcommand{\baselinestretch}{1.25}

\newcommand{\beeq}{\begin{equation}}
\newcommand{\eneq}{\end{equation}}
\newcommand{\beqn}{\begin{eqnarray}}
\newcommand{\eeqn}{\end{eqnarray}}

\def\mybig{\displaystyle \strut }

\def\dd{\partial}
\def\la{\raise.16ex\hbox{$\langle$}\lower.16ex\hbox{}  }
\def\ra{\, \raise.16ex\hbox{$\rangle$}\lower.16ex\hbox{} }
\def\go{\rightarrow}

\def\onehalf{ \hbox{${1\over 2}$} }

\def\Tr{{\rm Tr \,}}

\def\eff{{\rm eff}}
\def\min{{\rm min}}

\def\min{{\rm min}}

\def\psibar{ \psi \kern-.65em\raise.6em\hbox{$-$} }
\def\psibarl{ \psi \kern-.65em\raise.6em\hbox{$-$} \lower.6em\hbox{} }

\def\myfrac#1#2{{\mybig #1\over \mybig #2}}

\begin{document}
\thispagestyle{empty}

\baselineskip=12pt

{\small \noindent \mydate    \hfill OU-HET 519/2005}
\leftline{\small (Corrected 4-14-05)}


\baselineskip=35pt plus 1pt minus 1pt

\vskip 3.cm

\begin{center}
{\Large \bf Higgs Boson Mass and Electroweak-Gravity Hierarchy}\\
{\Large \bf from Dynamical Gauge-Higgs Unification}\\
{\Large \bf in the Warped Spacetime}\\


\vspace{2.5cm}
\baselineskip=20pt plus 1pt minus 1pt

{\bf  Yutaka Hosotani\footnote{hosotani@het.phys.sci.osaka-u.ac.jp} 
and Mitsuru Mabe\footnote{mitsuru@het.phys.sci.osaka-u.ac.jp} }\\
\vspace{.3cm}
{\small \it Department of Physics, Osaka University,
Toyonaka, Osaka 560-0043, Japan}\\
\end{center}

\vskip 2.3cm
\baselineskip=20pt plus 1pt minus 1pt

\begin{abstract}
Dynamical electroweak symmetry breaking by the Hosotani mechanism 
in the Randall-Sundrum warped spacetime is examined,  relations
among the W-boson mass ($m_W$), the Kaluza-Klein  mass scale ($M_{KK}$), 
and the Higgs boson mass ($m_H$)  being derived. It is shown that 
$M_{KK}/m_W \sim (2\pi kR)^{1/2} (\pi/\theta_W)$ and
$m_H /m_W \sim 0.058 \cdot kR (\pi/\theta_W)$, where
$k^2$, $R$, and $\theta_W$ are the curvature and size of the 
extra-dimensional space and  the Wilson line phase 
determined dynamically.  
For typical values  $kR = 12$ and $\theta_W = (0.2 \sim 0.4) \, \pi $, 
one finds that
$M_{KK}  = (1.7 \sim 3.5)$~TeV, 
$k = (1.3 \sim 2.6) \times 10^{19}$~GeV,  
and  $m_H = (140 \sim 280)$~GeV.
\end{abstract}

\newpage


\newpage

Although the standard model of the electroweak interactions has been 
successful to account for all the experimental data so far observed, 
there remain a few major issues to be settled.  First of all, Higgs particles 
are yet to be discovered. The Higgs sector of the standard model is
for the most part unconstrained unlike the gauge sector where the gauge principle 
regulates the interactions among matter.  Secondly, the origin of 
the scale of the electroweak interactions characterized by the W-boson 
mass $m_W \sim 80 \,$GeV or the vacuum expectation value of the Higgs 
field $v \sim 246 \,$GeV becomes mysterious once one tries to unify
the electroweak interactions with the strong interactions in the 
framework of grand unified theory, or with gravity, where the
energy scale is given by $M_{\rm GUT} \sim 10^{15} \,$-$\, 10^{17} \,$GeV or
$M_{\rm Pl} \sim 10^{19} \, $GeV, respectively. The natural explanation
of such hierarcy in the energy scales is desirable.  In this paper
we show that the Higgs sector of the electroweak interactions 
can be integrated in the gauge sector, and
the electroweak energy scale is naturally placed with the gravity scale
within the framework of dynamical gauge-Higgs unification in the 
Randall-Sundrum warped spacetime.

The scheme of dynamical gauge-Higgs unification was put forward long time 
ago in the context of higher dimensional non-Abelian gauge theory 
with non-simply connected extra-dimensional space.\cite{YH1,YH2}   In non-simply 
connected space there appear non-Abelian Aharonov-Bohm phases, or 
Wilson line phases, which can dynamically induce gauge symmetry breaking
even within configurations of vanishing field strengths.  The
extra-dimensional components of gauge potentials play a role of 
 Higgs fields
in four dimensions.  The Higgs fields are unified with the gauge fields 
and the gauge symmetry is dynamically broken at the quantum level.
It was originally designed that  Higgs fields in the adjoint 
representation  in $SU(5)$ grand unified theory are unified with 
the gauge fields.
  
The attempt to identify scalar fields as parts of gauge fields was
made earlier by utilizing symmetry reduction.  Witten observed that 
 gauge theory in four-dimensional Minkowski spacetime 
 with spherical symmetry reduces to a 
system of gauge fields and scalar fields in two-dimensional curved 
spacetime.\cite{Witten1}  This idea was extended to six-dimensional gauge theory
by Fairlie\cite{Fairlie1} and by Forgacs and Manton\cite{Manton1}
 to accommodate the electroweak
theory in four dimensions.  It was recognized there that to yield
$SU(2)_L \times U(1)_Y$ symmetry of electroweak interactions in
four dimensions one need start with a larger gauge group such as
$SU(3)$, $SO(5)$ or $G_2$.  The reduction of the symmetry to 
$SU(2)_L \times U(1)_Y$ was made by an ad hoc ansatz for field
configurations in the extra-dimensional space.  For instance, Manton 
assumed spherically symmetric configurations in the extra-dimensional 
space $S^2$. As was pointed out later,\cite{YH4} such a configuration 
can be realized  by a monopole configuration on $S^2$.\footnote{The monopole 
configuration for $A_M^8$ of the $SU(3)$ gauge fields on $S^2$ realizes
the envisaged symmetry reduction to $SU(2) \times U(1)$  in ref.\
\cite{Manton1}.}  
However, classical non-vanishing 
field strengths in  the background would lead to the 
instability of the system. In this regard gauge theory defined 
on non-simply  connected spacetime has big advantage in the sense that 
even with vanishing field strengths Wilson line phases become 
dynamical and can induce symmetry breaking at the quantum level 
by the Hosotani mechanism.  

Recently significant progress has been achieved along this line by
considering gauge theory on orbifolds which are obtained by modding
out non-simply connected space by discrete symmetry such as 
$Z_n$.\cite{Pomarol1}-\cite{YH3}
With the orbifold symmetry breaking induced from boundary 
conditions at fixed points of the orbifold,
a part of light modes in the Kaluza-Klein tower expansion of 
fields are eliminated from the spectrum at low energies so that chiral 
fermions in four dimensions  naturally emerge.\cite{Pomarol1}  
Further, in $SU(5)$ grand unified theory  (GUT)
on orbifolds  the triplet-doublet mass splitting problem of the Higgs 
fields \cite{Kawamura} and the gauge hierarchy problem\cite{Lim2}
can be naturally solved. 

The orbifold symmetry breaking, however, accompanies indeterminacy 
in theory.  It poses the arbitrariness problem  of boundary 
conditions.\cite{YH5}  One needs to show how and why a particular
set of boundary conditions is chosen naturally or dynamically,
which is achieved, though partially,  in the scheme of dynamical gauge-Higgs 
unification.

Quantum dynamics of Wilson line phases in GUT on orbifolds was 
first examined in ref.\ \cite{HHHK} where it was shown that the physical
symmetry is determined by the matter content.  Several attempts to
implement dynamical gauge-Higgs unification in the electroweak theory
have been made since then.  The most intriguing among those is the 
$U(3) \times U(3)$ model of Antoniadis, Benakli and Quiros.\cite{Antoniadis1}  
The effective potential of the Wilson line phases in this model has
been recently evaluated to show that the electroweak symmetry
breaking dynamically takes place with minimal addition of heavy
fermions.\cite{HNT2}  The model is restrictive enough to predict the 
Kaluza-Klein mass scale ($M_{KK}$) and the Higgs boson mass ($m_H$) 
with the $W$-bosn mass ($m_W$) as an input.  It turned out that 
$M_{KK} \sim 10 \, m_W$ and $m_H \sim \sqrt{\alpha_w}\,  m_W$, 
which contradicts with the observation.

We argue that this is not a feature of the specific model examined,
but is a general feature of orbifold models in which 
extra-dimensional space is flat.  Unless tuning of matter content 
is enforced, the relation $m_H \sim \sqrt{\alpha_w} \, m_W$ is 
unavoidable in flat space as shown below.  To circumvent this 
difficulty, it is necessary to have curved extra-dimensional space.

Randall and Sundrum introduced  warped spacetime 
with an extra-dimensional space having topology of $S^1/Z_2$
which is five-dimenional  anti-de Sitter spacetime with 
boundaries of two flat four-dimensional branes.\cite{RS1}  
It was argued there
that the standard model of electroweak interactions is placed on one of 
the branes such that the electroweak scale becomes natural compared
with the Planck scale chracterizing gravity.
Since then many variations of the Randall-Sundrum model have 
been investigated.  The standard model can be placed in the bulk
five-dimensional spacetime, not being restricted on one of the 
branes.\cite{Chang}  However, fine-tuning of the Higgs potential 
remains necessary.

More promising is to consider dynamical gauge-Higgs unification
in the Randall-Sundrum background where gauge theory is defined 
in the bulk five-dimensional spacetime without five-dimensional
scalar fields. The first step in this direction
has been made by Oda and Weiler who evaluated the 1-loop effective
potential for  Wilson line phases in the $SU(N)$  gauge theory.\cite{Oda1}
We will show in the present paper that the electroweak symmetry
breaking can be naturally implemented in dynamical gauge-Higgs
unification on the Randall-Sundrum background to avoid the
aforementioned difficulty concerning $M_{KK}$ and $m_H$.  
We show that in this scheme the Higgs mass $m_H$ should be between 
140 GeV and 280 GeV, and the Kaluza-Klein mass scale $M_{KK}$
must be between 1.7 TeV and 3.5 TeV.  It is exciting that
the predicted ranges of $m_H$ and $M_{KK}$ fall in the region 
where experiments at LHC can explore in the near 
future.\footnote{Cosmological consequences of the Hosotani 
mechanism in curved spacetime has been previously investigated in ref.\
\cite{HoNg}.  The Hosotani mechanism in the  Randall-Sundrum warped 
spacetime has been applied to the electroweak symmetry breaking 
 in ref.\ \cite{Pomarol2}.}

We consider  gauge theory 
in the Randall-Sundrum warped spacetime whose metric is given by
\beeq
ds^2 = e^{-2\sigma(y)} \eta_{\mu\nu} dx^\mu dx^\nu + dy^2 ~, 
\label{RSmetric1}
\eneq
$(\mu, \nu = 0,1,2,3)$.
Here $\sigma(y) = k |y|$ for $|y| \le \pi R$, 
$\sigma(y + 2\pi R) = \sigma(y)$ and 
$\eta_{\mu\nu} = {\rm diag~} (-1, 1,1,1)$.  Points $(x^\mu, -y)$ and
$(x^\mu, y+ 2\pi R)$ are identified with $(x^\mu, y)$. The resultant spacetime 
is an anti-de Sitter spacetime ($0< y < \pi R$) sandwitched
by  four-dimensional spacetime branes at $y=0$ and $y=\pi R$.
It has topology of $R^4 \times (S^1/Z_2)$. The curvature is given by $k^2$. 

As a prototype of the models we take
the $U(3)_S\times U(3)_W$ gauge theory\cite{Antoniadis1}, 
though the results do not depend on the details of the model.
Weak $W$ bosons reside in the $U(3)_W$ gauge group.
The $U(3)_W$ part of the action  is 
$I = \int d^5x \, \sqrt{- \det g} \, \big\{ -\onehalf \Tr F_{MN} F^{MN} +
{\cal L}_{\rm matter} \big\}$
where the five-dimensional coordinates are $x^M = (x^\mu, y)$
and ${\cal L}_{\rm matter}$ represents the part for quarks and leptons.
Five-dimensional scalar fields are not introduced.  The zero modes of the 
extra-dimensional components of the vector potentials, $A_y^a$, 
generate non-Abelian Aharonov-Bohm phases (Wilson line phases) 
and serve as four-dimensional Higgs fields effectively.

To see it more clearly, it is convenient to work in a new coordinate system 
$x^M = (x^\mu, w)$ where $w = e^{2ky}$ for $0 \le y \le \pi R$.  The metric 
becomes
\beeq
ds^2 = {1\over w} \, \eta_{\mu\nu} dx^\mu dx^\nu + 
{1\over 4k^2 w^2} \, dw^2 ~. 
\label{RSmetric2}
\eneq
Boundary conditions for the gauge potentials in the original
coordinate system $(x^\mu, y)$ are given in the form 
$(A_\mu, A_y)(x, y_j - y) = P_j (A_\mu, -A_y) (x, y_j + y) P_j^\dagger$
where $y_0 =0$, $y_1= \pi R$,  $P_j \in U(3)$ and $P_j^2 = 1$ 
($j=0,1$).\cite{HHHK,HNT1,HNT2} 
They follow from  the $S^1/Z_2$ nature of the spacetime.  In the
new coordinate system   $(x^\mu, w)$, the boundary conditions are
summarized as
\beqn
&&\hskip -1cm
 \pmatrix{A_\mu \cr A_w} (x, w_j)
 = P_j \pmatrix{A_\mu \cr - A_w} (x, w_j) P_j^\dagger ~~, \cr
 \noalign{\kern 5pt}
&&\hskip -1cm
\pmatrix{\dd_w A_\mu \cr \dd_w A_w} (x, w_j)
 = P_j \pmatrix{- \dd_w A_\mu \cr  \dd_w A_w} (x, w_j)  P_j^\dagger ~~,
\label{BC1}
\eeqn
where $w_0=1$ and $w_1=e^{2\pi kR}$.  Similarly, for a fermion in the 
fundamental representation 
\beqn
&&\hskip -1cm
\psi (x, w_j)  = \eta_j P_j \gamma^5 \psi(x, w_j) ~~, \cr
\noalign{\kern 5pt}
&&\hskip -1cm
 \dd_w \psi (x, w_j)  = -\eta_j P_j\gamma^5 \dd_w \psi(x, w_j) ~~,
\label{BC2}
\eeqn
where $\eta_j = \pm 1$. We take
\beeq
P_0 = P_1 = \pmatrix{-1 \cr & -1\cr && 1} 
\label{BC3}
\eneq
to ensure the electroweak symmetry.

The advantage of the $w$ coordinate over the $y$ coordinate lies in the
fact that zero modes of $A_w(x,w)$ become independent of $w$.  In the 
$y$ coordinate $A_y(x,y)$ has cusp singularities at $y=0$ and $y=\pi R$.
To observe it explicitly, we specify the gauge-fixing term in 
the action.  A general procedure in curved spacetime has been given in
ref.\ \cite{YH2}.  It is  convenient to adopt the prescription for 
gauge-fixing given in ref.\ \cite{Oda1}.  As is justified a posteriori,
the effective potential is evaluated in the background field method
with a constant background $A_M^c = \delta_{Mw} A_w^c$.  The gauge 
fixing term  $\int d^4x dw  \sqrt{-g} ~ {\cal L}_{\rm g.f.}$ is
chosen to be 
\beeq
{\cal L}_{\rm g.f.} 
= - w^2 \, \Tr (D_{\mu}^c A^{\mu}+4k^2 w D_w^c A_w)^2
\label{GaugeFix1}
\eneq
where $D_M^c A_N \equiv \dd_M A_N + ig [A_M^c, A_N]$ and 
$D_{\mu}^c A^{\mu} \equiv \eta^{\mu\nu} D_{\mu}^c A_{\nu}$.  In the path 
integral formula we write $A_M = A_M^c + A_M^q$ and expand the action 
in $A_M^q$.  The bilinear part of the action including the ghost part is 
given by
\beqn
&&\hskip -1cm 
I_\eff =-\int d^4 x\int_{w_0}^{w_1} dw \,
\bigg\{
\frac{1}{2kw} ~\Tr A_{\nu}^q (\dd^{\mu}\dd_{\mu} +4k^{2}w D_w^c D_w^c) A^{q\nu} \cr
\noalign{\kern 10pt}
&&\hskip 3.2cm 
+ 2k ~ \Tr A_w^q (\dd^{\mu}\dd_{\mu} +4k^2  D_w^c w D_w^c) A_w^q \cr
\noalign{\kern 10pt}
&&\hskip 3.2cm 
- {1\over 2kw^2} ~ 
\Tr \bar \eta \, (\dd^{\mu}\dd_{\mu} +4k^{2}w D_w^c D_w^c) \, \eta ~ \bigg\} ~.
\label{action2}
\eeqn
Partial integration necessary in deriving (\ref{action2}) is justified 
as $\Tr A_\mu \dd_w A^\mu$ and $\Tr A_w \dd_w A_w$ vanish at $w=w_0, w_1$
with the boundary conditions (\ref{BC1}).

Let us denote $A_M = \sum_{a=0}^8 \onehalf \lambda^a A^a_M$ with the standard
Gell-Mann matrices $\lambda^a$. ($\lambda^0$  represents the 
$U(1)$ part.)  With (\ref{BC1}) and (\ref{BC3}),   $A_\mu^a$ ($a=0,1,2,3,8$) and
$A_w^b$ ($b=4,5,6,7$) satisfy Neumann boundary conditions at $w=w_0, w_1$, 
whereas $A_\mu^a$ ($a=4,5,6,7$) and
$A_w^b$ ($b=0,1,2,3,8$) satisfy  Dirichlet boundary conditions.  
Zero modes independent of $w$ are allowed for  $A_\mu^a$ ($a=0,1,2,3,8$) and
$A_w^b$ ($b=4,5,6,7$).  It is found from (\ref{action2}) that they indeed
constitute massless particles in four dimensions when $A_w^c=0$. Gauge fields of
$SU(2)_L \times U(1)_Y$ are in $A_\mu^a$ ($a=0,1,2,3,8$), whereas
doublet Higgs fields are in $A_w^b$ ($b=4,5,6,7$).  We note that in the 
$y$ coordinate system $A_y'  = 2k e^{2ky} A_w$ so that the zero modes are not
constant in $y$, which gives rise to unphysical cusp singularities at 
$y=0, \pi R$.

Mode expansion for $A_\mu(x,w)$ is infered from (\ref{action2}) to be
\beqn
&&\hskip -1cm
A_\mu^a(x,w)=\sum_n A_{\mu,n}^a (x) f_n(w) ~, \cr
\noalign{\kern 10pt}
&&\hskip -1cm
-4 k^2 w \frac{d^2}{dw^2} f_n(w) =\lambda_n f_n(w) ~, \cr
\noalign{\kern 10pt}
&&\hskip -1cm
\int_{w_0}^{w_1}dw ~  \frac{1}{2kw} f_n(w) f_m(w) =\delta_{nm}  ~~.
\label{mode1}
\eeqn
For $A_w (x,w)$ one finds
\beqn
&&\hskip -1cm
A_w^a(x,w)=\sum_n A_{w,n}^a(x) h_n(w) ~, \cr
\noalign{\kern 10pt}
&&\hskip -1cm
-4k^2  \frac{d}{dw} w \frac{d}{dw} h_n(w) ={\hat{\lambda}}_n h_n(w) ~, \cr
\noalign{\kern 10pt}
&&\hskip -1cm
\int_{w_0}^{w_1}dw ~ 2k h_n(w) h_m(w) = \delta_{nm} ~~.
\label{mode2}
\eeqn
Given boundary conditions, $(\lambda_n, f_n(w))$ and $(\hat \lambda_n, h_n(w))$
are determined.  $A_\mu^a$ ($A_w^a$) has a zero mode $\lambda_0 = 0$ 
($\hat \lambda_0 = 0$) only with Neumann boundary conditions
at  $w=w_j$.  For the zero modes $f_0(w) = 1/\sqrt{\pi R}$ and
$h_0(w) = 1/\sqrt{2k(w_1 - w_0)}$. \cite{Lim3}

Except for the zero modes, both $\lambda_n$ and $ \hat \lambda_n $
are positive.  Apart from the normalization factors eigen functions are 
given by $f_n(w) = \sqrt{w} Z_1(\sqrt{\lambda_n w}/k)$ and 
$h_n(w) = Z_0(\sqrt{\hat \lambda_n w}/k)$ where $Z_\nu (z)$ is a
linear combination of Bessel functions $J_\nu(z)$ and $Y_\nu(z)$ of 
order $\nu$.  $(\lambda_n, f_n)$ with the Neumann boundary conditions
and  $(\hat \lambda_n, h_n)$ with the Dirichlet boundary conditions
are determined by
\beeq
\frac{J_0(\beta_n \sqrt{w_0})}{Y_0(\beta_n \sqrt{w_0})}=
\frac{J_0(\beta_n \sqrt{w_1})}{Y_0(\beta_n \sqrt{w_1})} ~~,
\label{eigen1}
\eneq
whereas $(\lambda_n, f_n)$ with the Dirichlet boundary conditions
and  $(\hat \lambda_n, h_n)$ with the Neumann boundary conditions
are determined by
\beeq
\frac{J_1(\beta_n \sqrt{w_0})}{Y_1(\beta_n \sqrt{w_0})}=
\frac{J_1(\beta_n \sqrt{w_1})}{Y_1(\beta_n \sqrt{w_1})} ~~.
\label{eigen2}
\eneq
Here $\beta_n = \sqrt{\lambda_n}/k$ or $\sqrt{\hat \lambda_n}/k$.
For $\beta_n \gg 1$, $\beta_n = \pi n/(\sqrt{w_1} - \sqrt{w_0})$. 
For $w_1^{-1/2} \ll \beta_n \ll 1$, 
$\beta_n = (n - {1\over 4})\pi/\sqrt{w_1}$ 
or $(n + {1\over 4})\pi /\sqrt{w_1}$ for
the case (\ref{eigen1}) or (\ref{eigen2}), respectively.
The first excited state is given by $\beta_1 \sqrt{w_1} \sim 2.6$ or $3.8$.
Hence, the Kaluza-Klein mass scale is given by
\beeq
M_{KK} = \myfrac{\pi k}{\sqrt{w_1} - \sqrt{w_0}} 
= \cases{R^{-1} &for $k \go 0$,\cr 
\noalign{\kern 5pt}
\pi k e^{-\pi kR} &for $e^{\pi kR} \gg 1$. \cr}
\label{KK1}
\eneq

With $P_j$  in (\ref{BC3}), the W boson and the weak Higgs doublet $\Phi$
are contained in the zero modes of $(A_\mu^1 \pm i A_\mu^2)(x,w)$ 
and $A_w^b(x,w)$  ($b=4,5,6,7$);
\beqn
&&\hskip -1cm 
{1\over \sqrt{2}}(A_\mu^1 + i A_\mu^2)(x,w)
\Rightarrow
{1\over \sqrt{2}}(A_{\mu,0}^1 + i A_{\mu,0}^2)(x) f_0(w)
= {1\over \sqrt{\pi R}} W_\mu (x) ~, \cr
\noalign{\kern 10pt}
&&\hskip -1cm 
{1\over \sqrt{2}} \pmatrix{A_w^4 - i A_w^5 \cr A_w^6 - i A_w^7} (x,w)
\Rightarrow
{1\over \sqrt{2}}
  \pmatrix{A_{w,0}^4 - i A_{w,0}^5 \cr A_{w,0}^6 - i A_{w,0}^7} (x) h_0(w)
= {\Phi (x)\over \sqrt{2k(w_1 - w_0)}}  ~.
\label{W-Higgs1}
\eeqn
There is no potential term for $\Phi$ at the classical level,
but nontrivial effective potential is generated at the quantum level.
As in the model discussed in ref.\ \cite{HNT2}, the effective potential
is supposed to have a global minimum at $\Phi \not= 0$, inducing
dynamical electroweak symmetry breaking.  Making use of the residual
$SU(2) \times U(1)$ invariance, we need to evaluate the effective
potential for the configuration
\beeq
A_w = A_w^c = \alpha \Lambda ~~,~~
\Lambda = \pmatrix{\cr &&1 \cr & 1} ~~.
\label{phase1}
\eneq
Note that $v = \sqrt{2} \la \Phi^0 \ra = 2 \sqrt{2k(w_1 - w_0)} ~\alpha$.

The Randall-Sundrum warped spacetime has topology of $R^4 \times (S^1/Z_2)$.
As $S^1$ is not simply connected, there arise Aharonov-Bohm phases, or
Wilson line phases, which become physical degrees of freedom.\cite{YH1,YH2}  
 The Wilson line phases are defined by eigenvalues of
$P \exp \Big\{ ig \int_C dw A_w \Big\} \cdot U$ where the path $C$ is a 
closed non-contractible loop along $S^1$ and $U = P_1 P_0$.  
In the present case $U  = I$  so that all gauge potentials are 
periodic on $S^1$.  It follows that $\alpha$ in (\ref{phase1}) is
related to the Wilson line phase by
\beeq
\theta_W = 2 g \alpha (w_1 - w_0) ~~.
\label{phase2}
\eneq
It will be shown below that $\theta_W$ and $\theta_W + 2\pi$
are gauge equivalent.  The $SU(2)_L$ gauge coupling constant
in four dimensions, $g_4$, is easily found by inserting
$A_\mu(x,w) \sim (\pi R)^{-1/2} A_{\mu,0}(x)$ into $F_{\mu\nu}$;
\beeq
g_4 = \myfrac{g}{\sqrt{\pi R}} ~.
\label{coupling1}
\eneq

Nonvanishing $\theta_W$ or $v$ gives the $W$ boson a mass $m_W$.  
In our scheme the mass term for $W$ arises from the term
$- \int  dw  \, 2k \Tr A_\nu^q D_w^c D_w^c A^{q\nu}$ in 
(\ref{action2}).  The resultant relation is the standard one, 
$m_W = \onehalf g_4 v$.  Thus one finds
\beeq
m_W = \myfrac{g_4 v}{2} 
=  \left[\myfrac{\pi k}{2R(w_1 - w_0)}\right]^{1/2}
~ \myfrac{\theta_W}{\pi} 
= \cases{
\myfrac{1}{2}\myfrac{\theta_W}{\pi}M_{KK}  &for $k\go 0$~,\cr
 \myfrac{1}{\sqrt{2\pi k R}}
 \myfrac{\theta_W}{\pi} M_{KK}  &for $e^{\pi kR} \gg 1$~,\cr}
\label{WKK1}
\eneq
where $M_{KK}$ is given in (\ref{KK1}).

The precise value of $\theta_W$
depends on the details of the model.  If the effective potential is
minimized at $\theta_W=0$, then the electroweak symmetry breaking
does not occur.  If it occurs, $\theta_W$ takes a value typically 
around $0.2 \pi$ to $0.4 \pi$, unless artificial tuning of 
matter content is made.  As an example, in the model discussed
in ref.\ \cite{HNT2} in flat space, $\theta_W \sim 0.25 \pi$, 
which, with $m_W = 80.4\,$GeV inserted,  yielded
 too small $M_{KK} \sim 640 \,$GeV.  

In the present case, with the value of $\theta_W$ given, $kR$
determines $M_{KK}$ and $k$.  Recall that the four- and 
five-dimensional Planck constants $M_{\rm pl}$ and $M_{\rm 5d}$
are related by $M_{\rm pl}^2 k \sim  M_{\rm 5d}^3$.  To have
a natural relation $M_{\rm 5d} \sim M_{\rm pl}$, $kR$ must be
in the range $11 < kR <13$.  To confirm it, take $\theta_W = 0.25 \pi$
as an example.  For $kR = 12$, one finds $M_{KK} = 2.8\,$TeV and
$k = 2.1 \times 10^{19} \,$GeV.  However, for $kR = 6$ and $24$
one finds $k=9.7 \times 10^{10} \,$GeV and $7.0 \times 10^{35} \,$GeV,
respectively.  In (\ref{WKK1}), $M_{KK} /m_W \propto \sqrt{kR}$,
and $kR$ is about 12 if there is only one gravity 
scale ($M_{\rm 5d} \sim k$).  Thus the value of $M_{KK}$ is predicted to
be $1.7 \, {\rm TeV} < M_{KK} < 3.5 \, {\rm TeV}$ 
for $0.2 \pi < \theta_W < 0.4 \pi$ in the present scenario.

How about the Higgs boson mass?  The finite mass of the Higgs field $\Phi$ is 
generated by quantum effects.\cite{HHHK}  One needs to evaluate the effective 
potential for the Wilson line phase, $V_\eff (\theta_W)$.  The Higgs
mass is determined from the curvature at the minimum, with the 
substitution $\theta_W = g [(w_1 -w_0) k^{-1} \Phi^\dagger \Phi ]^{1/2}$.
Its magnitude is estimated reliably thanks to the phase nature
of $\theta_W$.

To prove that $\theta_W$ and $\theta_W + 2\pi$ are physically
equivalent, we go back to the boundary conditions (\ref{BC1}) and 
(\ref{BC2}) with general $P_j$. Let us perform a gauge transformation
$A_M' = \Omega A_M \Omega^\dagger - (i/g) \Omega \dd_M \Omega^\dagger$.
$A_M'$ does not satisfy the same boundary conditions as $A_M$ 
in general. Instead
\beqn
&&\hskip -1cm
 \pmatrix{A_\mu' \cr A_w'} (x, w_j)
 = P_j' \pmatrix{A_\mu' \cr - A_w'} (x, w_j) P_j'^\dagger ~~, \cr
\noalign{\kern 10pt}
&&\hskip -1cm
\pmatrix{\dd_w A_\mu' \cr \dd_w A_w'} (x, w_j)
 = P_j' \pmatrix{-\dd_w A_\mu' \cr  \dd_w A_w'} (x, w_j)  P_j'^\dagger 
 ~~,\cr
\noalign{\kern 10pt}
&&\hskip -1cm
P_j' = \Omega(x, w_j) P_j \Omega(x, w_j)^\dagger ~~, 
\label{BC4}
\eeqn
provided 
\beqn
&&\hskip -1cm
[P_j', \dd_\mu \Omega \Omega^\dagger(x, w_j)] 
= \{P_j' , \dd_w \Omega \Omega^\dagger(x, w_j)  \}  = 0 ~~, \cr
\noalign{\kern 10pt}
&&\hskip -1cm
\{P_j', \dd_\mu (\Omega \dd_w \Omega^\dagger)(x, w_j) \} 
= [P_j' , \dd_w (\Omega \dd_w \Omega^\dagger)(x, w_j) ]  = 0 ~~.
\label{BC5}
\eeqn
In general, $P_j'$ differs from $P_j$.  When the conditions in
(\ref{BC5}) are satisfied, the two sets of the boundary conditions
are said to be in the equivalence relation 
$\{ P_0, P_1 \} \sim \{ P_0', P_1' \}$, which defines equivalence 
classes of boundary conditions.  Extensive analysis of the equivalence
classes  of boundary conditions has been given in refs.\ 
\cite{YH2,HHHK,HHK}.  It was shown there that physics is the same
in each equivalence class of boundary conditions.

In the present context we are interested in the residual gauge
invariance which preserves the boundary conditions.  In particular
we would like to know $\Omega(x,w)$ which satisfies (\ref{BC5}) and
yields $P_j' = P_j$, but shifts $\theta_W$.
Take $( P_0, P_1)$ in (\ref{BC3}).  We perform a gauge transformation
\beeq
\Omega(x,w) = e^{i\beta (w-  w_0) \Lambda}
\label{gaugeT1}
\eneq
where $\Lambda$ is defined in (\ref{phase1}) and satisfies 
$\{ \Lambda , P_j \} = 0$.  Note that $P_0' = P_0$, 
$P_1' = e^{2i\beta (w_1 - w_0) \Lambda} P_1$,  and 
$\dd_w \Omega \Omega^\dagger = - \Omega \dd_w\Omega^\dagger 
=i \beta \Lambda$. All the conditions in (\ref{BC5}) are satisfied.
Further, for $\beta = n\pi/(w_1 - w_0)$ ($n$: an integer), 
$P_j' = P_j$, i.e.\ the boundary conditions are preserved.  For
the configuration $A_w$ in (\ref{phase1}), the new gauge potential is
$A_w' = (\alpha - [n\pi/g(w_1-w_0)]) \Lambda$.  
$\theta_W = 2g\alpha (w_1 - w_0)$ is shifted, under the gauge transformation
(\ref{gaugeT1}), to $\theta_W' = \theta_W - 2n\pi$.
$\theta_W$ and $\theta_W + 2\pi$ are related by a large gauge 
transformation so that they are physically equivalent.

Having established the phase nature of $\theta_W$, we estimate
$V_\eff(\theta_W)$. $V_\eff(\theta_W)$ in the models in flat 
orbifolds has been evaluated well.\cite{Lim1,HHHK,HHK,HNT1,HNT2}
$V_\eff(\theta_W)$ in the Randall-Sundrum
spacetime in the $SU(N)$ gauge theory has been evaluated by 
Oda and Weiler.\cite{Oda1}
With the background $A_w^c$ or $\theta_W$,
the spectrum $\lambda_n$ of each field degree of freedom depends
on $\theta_W$ as well as on the boundary conditions of the field.
Its contribution to four-dimensional $V_\eff(\theta_W)$ 
at the one loop level is summarized as
\beeq
V_\eff(\theta_W) = \mp {i\over 2} \int {d^4 p\over (2\pi)^4} \,
\sum_n \ln \Big\{ -p^2 + \lambda_n(\theta_W) \Big\}  ~~,
\label{effV1}
\eneq
where $-$ ($+$) sign is for a boson (fermion). The spectrum 
$\lambda_n$ for $\theta_W =0$ is determined as described in
the discussions from Eq.\ (\ref{action2}) to (\ref{eigen2}).
It is found there that $\lambda_n \sim M_{KK}^2 n^2$ for large $n$.
Hence one can write, after making a Wick rotation, as
\beeq
V_\eff(\theta_W) = \pm {1\over 2} M_{KK}^4 
 \int {d^4 q_E\over (2\pi)^4} \,
\sum_n \ln \Big\{ q_E^2 + \rho_n(\theta_W)\Big\} + {\rm constant} ~,
\label{effV2}
\eneq
where $\rho_n(\theta_W) = \lambda_n/M_{KK}^2$.  It is known that
on an orbifold with topology of $S^1/Z_2$, fields form a $Z_2$ 
doublet pair to have an interaction with $\theta_W$.\cite{HHHK}  The resultant
spectrum for a $Z_2$ doublet is cast in the form where the sum 
in (\ref{effV2}) extends over from $n=-\infty$ to $n=+\infty$.
Further $\rho_n(\theta_W + 2 \pi) = \rho_{n + \ell} (\theta_W)$ 
($\ell$: an integer), and 
$\rho_n(\theta_W) \sim [ n + \gamma(\theta_W)]^2$ for large $|n|$  where 
$\gamma(\theta_W+ 2\pi)=\gamma(\theta_W)+ \ell$.
For instance, in the $U(3) \times U(3)$ model in flat space,
$\rho_n(\theta_W) = [ n+ \ell \theta_W/2\pi + {\rm (const)}]^2$
with $\ell = 0, \pm 1, \pm 2$.\cite{HNT2}
The important feature is that as $\theta_W$ is shifted to 
$\theta_W + 2\pi$ by a large gauge transformation, each eigen mode
is shifted to the next KK mode in general, but the spectrum 
as a whole remains the same.

Recall the formula
\beqn
&&\hskip -1cm
{1\over 2} \int {d^4 q_E\over (2\pi)^4} \sum_{n=-\infty}^\infty
\ln \Big\{ q_E^2 + (n+x)^2 \Big\} 
= - {3\over 64 \pi^6} \, h(x) + {\rm constant} ~, \cr
\noalign{\kern 10pt}
&&\hskip 1cm
h(x) = \sum_{n=1}^\infty {\cos 2n \pi  x \over n^5} ~~.
\label{formula1}
\eeqn
The $x$-dependent part is finite. 
In the present case we have $ \sum (\pm) h[\gamma(\theta_W)]$.  The total
effective potential takes the form
\beeq
V_\eff(\theta_W) = N_\eff \,
{3\over 128\pi^6} M_{KK}^4 \, f(\theta_W) 
\label{effV3}
\eneq
where $f(\theta_W + 2\pi)= f(\theta_W)$ and its amplitude is 
normalized to be an unity.  Once the matter content of the model is specified,
the coefficient $N_\eff$ is determined.  In the minimal model
or its minimal extension, $N_\eff = O(1)$ as supported by examples.

When $V_\eff(\theta_W)$ has a global minimum at a nontrivial 
$\theta_W = \theta_W^{\rm min}$,  dynamical electroweak symmetry breaking 
takes place.  It typically happens at 
$\theta_W^{\rm min} = (0.2 \sim 0.3) \pi$.\cite{HNT2}
It is possible to have a very small $\theta_W^{\rm min} \sim 0.01 \pi$ by
fine-tuning of the matter content as shown in ref.\ \cite{HHKY}, which,
however,  is
eliminated in the present consideration for the artificial nature. 
The mass $m_H$  of the neutral Higgs boson is found by expanding 
$V_\eff(\theta_W)$ around $\theta_W^{\rm min}$ and using 
$\theta_W = g [(w_1 -w_0) k^{-1} \Phi^\dagger \Phi ]^{1/2}$.  One finds
\beeq
m_H^2 = N_\eff  f''(\theta_W^{\rm min}) 
\myfrac{3\alpha_w }{64 \pi^4 }  
\myfrac{R (w_1 - w_0)}{k} M_{KK}^4 
\label{HiggsMass1}
\eneq
where $\alpha_w = g_4^2/4\pi$. In a generic model $f''(\theta_W^\min) \sim 1$.
Making use of (\ref{KK1}) and (\ref{WKK1}), one finds
\beeq
m_H = \cases{c \, \bigg( \myfrac{3 \alpha_w}{32\pi^3}  \bigg)^{1/2}  M_{KK}
\hskip 0.9cm
= c \, \bigg( \myfrac{3 \alpha_w}{8\pi^3}  \bigg)^{1/2}
    \myfrac{\pi}{\theta_W^\min} \, m_W
&for $k \go 0$ ~, \cr
c \,  \bigg(\myfrac{3 \alpha_w}{64\pi^2} \bigg)^{1/2} \sqrt{kR} ~ M_{KK}
= c \, \bigg( \myfrac{3 \alpha_w}{32 \pi}  \bigg)^{1/2}
   kR  \myfrac{\pi}{\theta_W^\min} \, m_W
&for $e^{\pi kR} \gg 1$ ~, \cr}
\label{HiggsMass2}
\eneq
where $c= [N_\eff  f''(\theta_W^{\rm min})]^{1/2}$.  
The values of $\theta_W^\min$ and $c$ depend on details of the model.
In the models analysed in ref.\ \cite{HNT2},  $(\theta_W^\min ,c)$
ranges from $(0.269 \pi , 2.13)$ to $(0.224 \pi, 1.63)$, 
which justifies our estimate.
Hereafter we set $c=1.9$, understanding 20{\%} uncertainty.
Inserting $\alpha_w = 0.032$ and $kR = 12$, we obtain that 
$m_H = 0.70 (\pi/\theta_W^\min) m_W$ and  $M_{KK} = 12.4 m_H$.  
In flat space (in the $k \go 0$ limit),  
$m_H = 0.037 (\pi/\theta_W^\min) m_W$ and  $M_{KK} = 53.9 m_H$, 
which yielded too small $m_H$.  There appears a large enhancement factor $kR$
in the relation connecting $m_H$ and $m_W$ in the Randall-Sundrum warped 
spacetime.  For a typical value $\theta_W^\min = (0.2 \sim 0.4) \pi$, 
the mass of the Higgs boson and the Kaluza-Klein mass scale are 
 given by $m_H = (140 \sim 280) \,$GeV and $M_{KK} = ( 1.7 \sim 3.5) \,$TeV,
 respectively.

The relations (\ref{WKK1}) and (\ref{HiggsMass2}) reveal many remarkable facts.
First of all, only the parameter $kR$ in the Randall-Sundrum spacetime 
appears in the relations connecting $m_W$, $m_H$ and $M_{KK}$.  Secondly,
if one supposes that $k = O(M_{\rm pl})$, then $kR = 12 \pm 1$ to have
the observed value for $m_W$.  The electroweak-gravity hierarchy is accounted 
for  by a moderate value for $kR$. 
Thirdly, another quantity $\theta_W^\min$
involved in those relations  is dynamically determined, once the matter content
of the model is specified.  In case the electroweak symmetry breaking
takes place, it typically takes $(0.2 \sim 0.4) \pi$.  $m_H$ and $M_{KK}$
are predicted  up to the factor $\theta_W^\min$.  Fourthly and most
remarkably, the predicted value for $m_H$, $140 \sim 280 \,$GeV, is 
exactly in the range which can be explored in the experiments at LHC and
other planned facilities in the near future.  In conjunction with it,
we recall that in the minimal supersymmetric standard model
the Higgs boson mass is predicted in the range 
$100 < m_H <  130\,$GeV.\cite{Okada}  Experimentally preferred value is
$m_H = 126^{+73}_{-48} \,$GeV.\cite{HiggsExp}

In the dynamical gauge-Higgs unification the Higgs field in four 
dimensions is identified with the extra-dimensional component of the 
gauge fields.   The Hosotani mechanism induces dynamical electroweak 
symmetry breaking, giving both weak gauge bosons and Higgs boson finite 
masses.  The desirable enhancement factor for $m_H$ originates from the 
property that the Higgs field is a part of five-dimensional vector, not a 
scalar, whose coupling to gravity and matter differs from those of 
four-dimensional gauge fields in the Randall-Sundrum warped spacetime.

Our scenario significantly differs from the Higgsless model where 
four-dimensional Higgs fields are eliminated from the spectrum by ad hoc 
boundary conditions on orbifolds.\cite{Higgsless}
In our scenario there is a Higgs boson
with $m_H =(140 \sim 280)\,$GeV.  Its mass is generated by radiative corrections.  
There is no quadratic divergence associated with $m_H^2$ thanks to the 
gauge invariance in five dimensions.  As in supersymmetric theories 
the unitarity is expected to be assured by the existence of light Higgs boson.

The scenario of the dynamical gauge-Higgs unification in the warped
spacetime is promising.  In the present paper we focused on $m_H$ and
$M_{KK}$.  There are many issues to be examined.   Yukawa
couplings among fermions and the Higgs boson,   couplings of fermions
to Kaluza-Klein excitations of the gauge and Higgs bosons,
and self-couplings of the Higgs boson can be also explored
in the forthcoming experiments. It is also interesting to extend our
analysis to supersymmetric (SUSY) theories in the Randall-Sundrum 
spacetime.\cite{Takenaga1}
SUSY breaking scale $M_{\rm SUSY} \sim 1\,$TeV is not far from $M_{KK}$ 
in the present paper.   We shall come back to these issues 
in separate publications.

\vskip .5cm

\leftline{\bf Acknowledgments}
This work was supported in part by  Scientific Grants from the Ministry of 
Education and Science, Grant No.\ 13135215  and
Grant No.\ 15340078 (Y.H).



\vskip 1.cm

\def\jnl#1#2#3#4{{#1}{\bf #2} (#4) #3}

\def\Zphys{{\em Z.\ Phys.} }
\def\jssc{{\em J.\ Solid State Chem.\ }}
\def\jpsJ{{\em J.\ Phys.\ Soc.\ Japan }}
\def\ptps{{\em Prog.\ Theoret.\ Phys.\ Suppl.\ }}
\def\PTP{{\em Prog.\ Theoret.\ Phys.\  }}

\def\JMP{{\em J. Math.\ Phys.} }
\def\NPB{{\em Nucl.\ Phys.} B}
\def\NP{{\em Nucl.\ Phys.} }
\def\PLB{{\em Phys.\ Lett.} B}
\def\PL{{\em Phys.\ Lett.} }
\def\PRL{\em Phys.\ Rev.\ Lett. }
\def\PRB{{\em Phys.\ Rev.} B}
\def\PRD{{\em Phys.\ Rev.} D}
\def\PRe{{\em Phys.\ Rep.} }
\def\AP{{\em Ann.\ Phys.\ (N.Y.)} }
\def\RMP{{\em Rev.\ Mod.\ Phys.} }
\def\ZPC{{\em Z.\ Phys.} C}
\def\SCI{\em Science}
\def\CMP{\em Comm.\ Math.\ Phys. }
\def\MPLA{{\em Mod.\ Phys.\ Lett.} A}
\def\IJMPA{{\em Int.\ J.\ Mod.\ Phys.} A}
\def\IJMPB{{\em Int.\ J.\ Mod.\ Phys.} B}
\def\EPJC{{\em Eur.\ Phys.\ J.} C}
\def\PR{{\em Phys.\ Rev.} }
\def\JHEP{{\em JHEP} }
\def\cmp{{\em Com.\ Math.\ Phys.}}
\def\JPA{{\em J.\  Phys.} A}
\def\JPG{{\em J.\  Phys.} G}
\def\NJP{{\em New.\ J.\  Phys.} }
\def\CQG{\em Class.\ Quant.\ Grav. }
\def\ATMP{{\em Adv.\ Theoret.\ Math.\ Phys.} }
\def\ibid{{\em ibid.} }

\renewenvironment{thebibliography}[1]
         {\begin{list}{[$\,$\arabic{enumi}$\,$]}  
         {\usecounter{enumi}\setlength{\parsep}{0pt}
          \setlength{\itemsep}{0pt}  \renewcommand{\baselinestretch}{1.2}
          \settowidth
         {\labelwidth}{#1 ~ ~}\sloppy}}{\end{list}}

\def\reftitle#1{}                

\end{document}